\documentclass[preprint,12pt]{elsarticle}

\usepackage{graphicx}

\usepackage{amssymb}
\usepackage{graphicx}
\usepackage{subfigure}

\begin{document}

\begin{frontmatter}



\title{The Robustness of Scale-free Networks Under Edge Attacks with the Quantitative Analysis}
\author[1,2,3]{Bojin ZHENG\corref{cor1}}
\ead{zhengbojin@gmail.com}
\author[1]{Hongrun WU}
\author[1]{Wenhua DU\corref{cor2}}
\ead{dr.duwh@gmail.com}
\author[1]{Wanneng SHU}
\author[1,2]{Jun QIN}

\address[1]{College of Computer Science, South-Central University for Nationalities, Wuhan 430074, China}
\address[2]{State Key Laboratory of Networking and Switching Technology, Beijing University of Posts and  Telecommunications, Beijing 100876, China}
\address[3]{School of Software, Tsinghua University, Beijing 100084,China}
\begin{abstract}
Previous studies on the invulnerability of scale-free networks under edge attacks supported the conclusion that scale-free networks would be fragile under selective attacks. However, these studies are based on qualitative methods with obscure definitions on the robustness. This paper therefore employs a quantitative method to analyze the invulnerability of the scale-free networks, and uses four scale-free networks as the experimental group and four random networks as the control group. The experimental results show that some scale-free networks are robust under selective edge attacks, different to previous studies. Thus, this paper analyzes the difference between the experimental results and previous studies, and suggests reasonable explanations.
\end{abstract}


\end{frontmatter}



\section{Introduction}

Previous studies on the invulnerability of scale-free networks suggested that the robustness and fragility coexist, i.e., scale-free networks are robust under random errors yet fragile under selective node attacks\cite{reka138}. Later, Holme et al. explored the invulnerability under the edge attacks\cite{a18}, and concluded that scale-free networks are still fragile under selective edge attacks, but much more robust than under selective node attacks.

However, previous studies are based upon an obscure definition of the robustness, i.e., the rapid decay of network performance curves. Unfortunately, even the complete networks, which are the most robust, can not be thought robust, because their network performance curves would also rapidly decay under the node attacks. Therefore, it is worth of arguing when judging the robustness by the decay rate of network performance curves.

This paper utilizes a quantitative method\cite{jun220}, which strictly defines the robustness and fragility, to reinvestigate the robustness of scale-free networks under edge attacks. Interestingly, the experimental results show that some scale-free networks are fragile under selective edge attacks, yet some are unexpected robust. The experimental results are different to previous conclusions.

To solve the difference, this paper suggests reasonable explanations to make the experimental results and previous conclusions consistent.

\section{Results}

Previous studies on the invulnerability of scale-free networks drawn a conclusion that scale-free networks are fragile under selective edge attack. Because the experiments in these studies commonly utilized the networks from the real world or the theoretical models and did not set the control group, this kind of case studies can not distinguish the impact of different possible factors, therefore, we can not attribute the robustness or fragility to the scale-free property.

Zheng et al.'s theory\cite{152} has demonstrated that the average degree may be a factor to the invulnerability of scale-free networks under selective node attacks. Therefore, the robustness or fragility of scale-free networks under selective edge attacks probably have alterative explanations. To eliminate the impacts of the relative factors, we should set the control group so that we can focus on the impact of the scale-free property. Therefore, we choose four scale-free networks as the experimental group, and generate a random network for each of the selected scale-free networks and use these random networks as the control group. Every pair of networks have the same node number and edge number. By this means, we can eliminate the impact of different average degrees.

Previous studies did not define the robustness quantitatively, therefore, the corresponding conclusions are obscure and arguable. This paper employs the $I$ index \cite{jun220} to deal with this problem. The employed index defines a threshold value 0 to distinguish the robustness and fragility, and it can deal with both node and edge attacks and assure the similar results.

Similar to most previous studies, this paper uses three typical attack strategies to explore the robustness of the scale-free networks.

\subsection{The Selected Networks}
We select four scale-free networks as the experimental group, which are the CSF compact network\cite{152}(denoted as CSF), the CSF non-compact network (denoted as CSFN), the political book network (denoted as Polbook\footnote{This network was edited by Valdis Krebs and was downloaded from MEJ Newman's website.}) and the protein network\cite{a13}(denoted as Protein).  The CSF has a center with dense links. The CSFN and the Polbook network have community structures. As to the average degree value (denoted as $\overline{D}$), the CSF and the CSFN are larger, the Polbook is medium and the Protein is smaller.

We generate four random networks as the control group accordingly to four scale-free networks in the experimental group. Besides, for each random network, we use the name of the corresponding scale-free network with a prime symbol to denote it.

We list the properties of the selected networks and corresponding random networks as Table \ref{tab1}.

\begin{table}[htbp]\caption{The properties of networks.}\label{tab1}
\centering
\begin{tabular}
{|p{50pt}|p{40pt}|p{40pt}|p{40pt}|p{65pt}|}
\hline
Network&
Node&
Edge&
$\overline{D}$&
Compactness \\
\hline
CSF&
1200&
7079&
11.798&
0.943 \\
\hline
CSF'&
1200&
7079&
11.798&
- \\
\hline
CSFN&
1200&
6837&
11.395&
0.394 \\
\hline
CSFN'&
1200&
6837&
11.395&
- \\
\hline
Polbook&
105&
441&
8.4&
0.6971 \\
\hline
Polbook'&
105&
441&
8.4&
- \\
\hline
Protein&
1458&
1948&
2.67&
0.8463 \\
\hline
Protein'&
1458&
1948&
2.67&
- \\
\hline
\end{tabular}
\end{table}

\subsection{The Invulnerability Index}

Similar to Schneider's $R$ index\cite{onion,318}, the $I$ index also uses the area to measure the invulnerability, but the $I$ index defines a baseline and uses the area above the baseline (meaning robustness) minus the area under the baseline (meaning fragility) to measure the invulnerability. Moreover, the $I$ index sums the fractions of edges instead of nodes in the largest connected cluster.

Because the node attacks can be transformed into the edge attacks, it is very important to assure that the value of the index under the node attacks approximates to the value under the edge attacks. Because the $I$ index is based on the fractions of edges, it easily holds such a feature.

\subsection{The Edge Attacks Strategies}

We choose three commonly used edge attack strategies to carry out the experiments, i.e., the random edge attack (RnE), the initial-graph degree-based edge attack (IDE) and the initial-graph betweenness-based edge attack (IBE). RnE means that all edges are treated equally and chosen randomly. IDE means that the edges would be removed one by one according to the highest edge degrees which are calculated from the initial networks. IBE is similar to IDE, but according to the edge betweenness instead of the edge degrees.

We define the edge degree $D_E$ of the edge $e$ as the product of power of the degrees of two ends which is expressed as equation \ref{eq1}.

\begin{equation}\label{eq1}
    D_E(e) = x_i^{\varpi}x_j^{\varpi}
\end{equation}

where $x_i$ and $x_j$ are the degree value of the nodes in two ends of the edge $e$.
To carry out the experiments, this paper sets $\varpi$ = $1$.

\subsection{The Experimental Results}

\begin{figure}[htbp]
  \subfigure[CSF]{
    \label{fig:subfig:a} 
    \includegraphics[width=3.6cm,height=3cm]{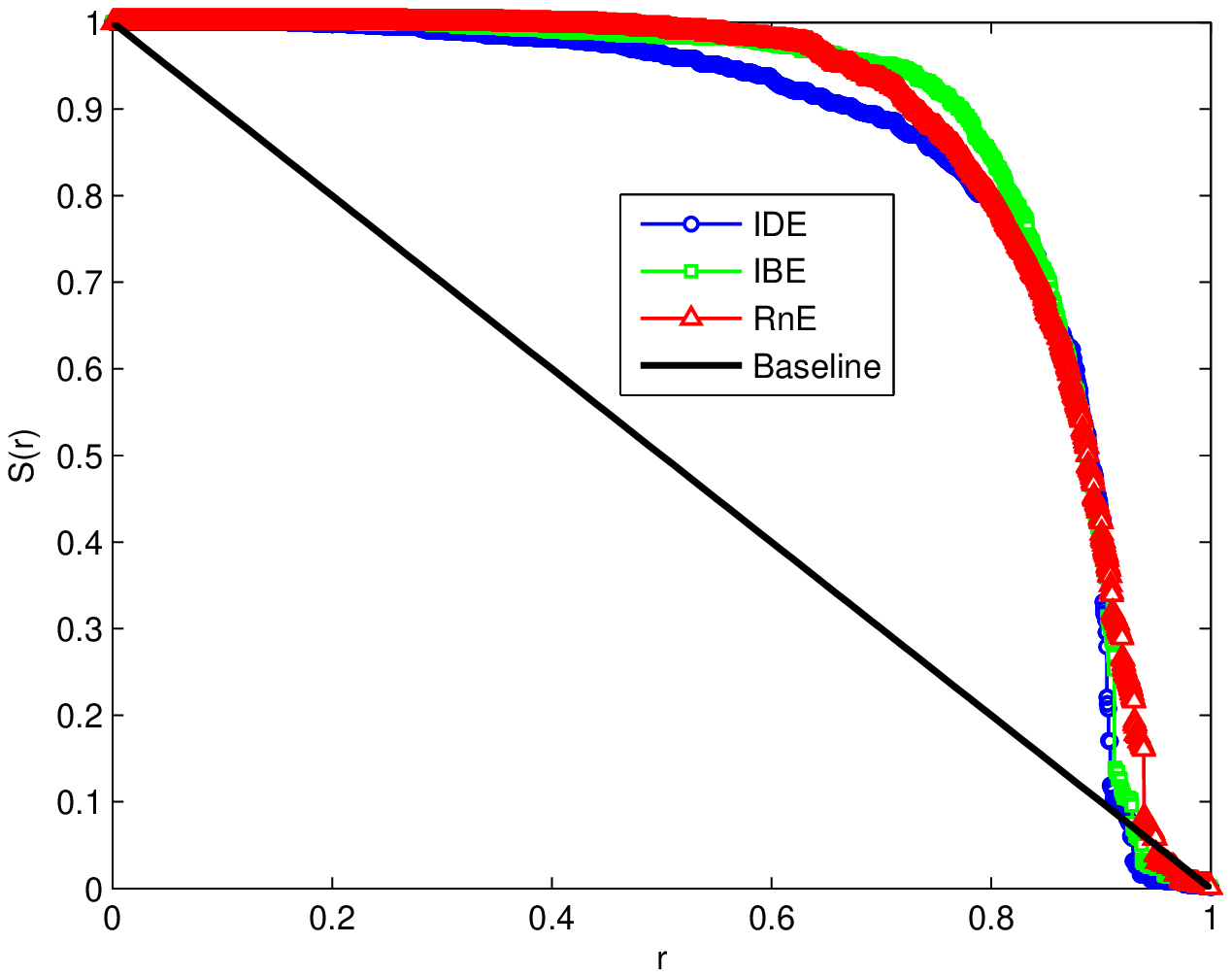}}
  \hspace{-0.7cm}
  \subfigure[CSFN]{
    \label{fig:subfig:b} 
    \includegraphics[width=3.6cm,height=3cm]{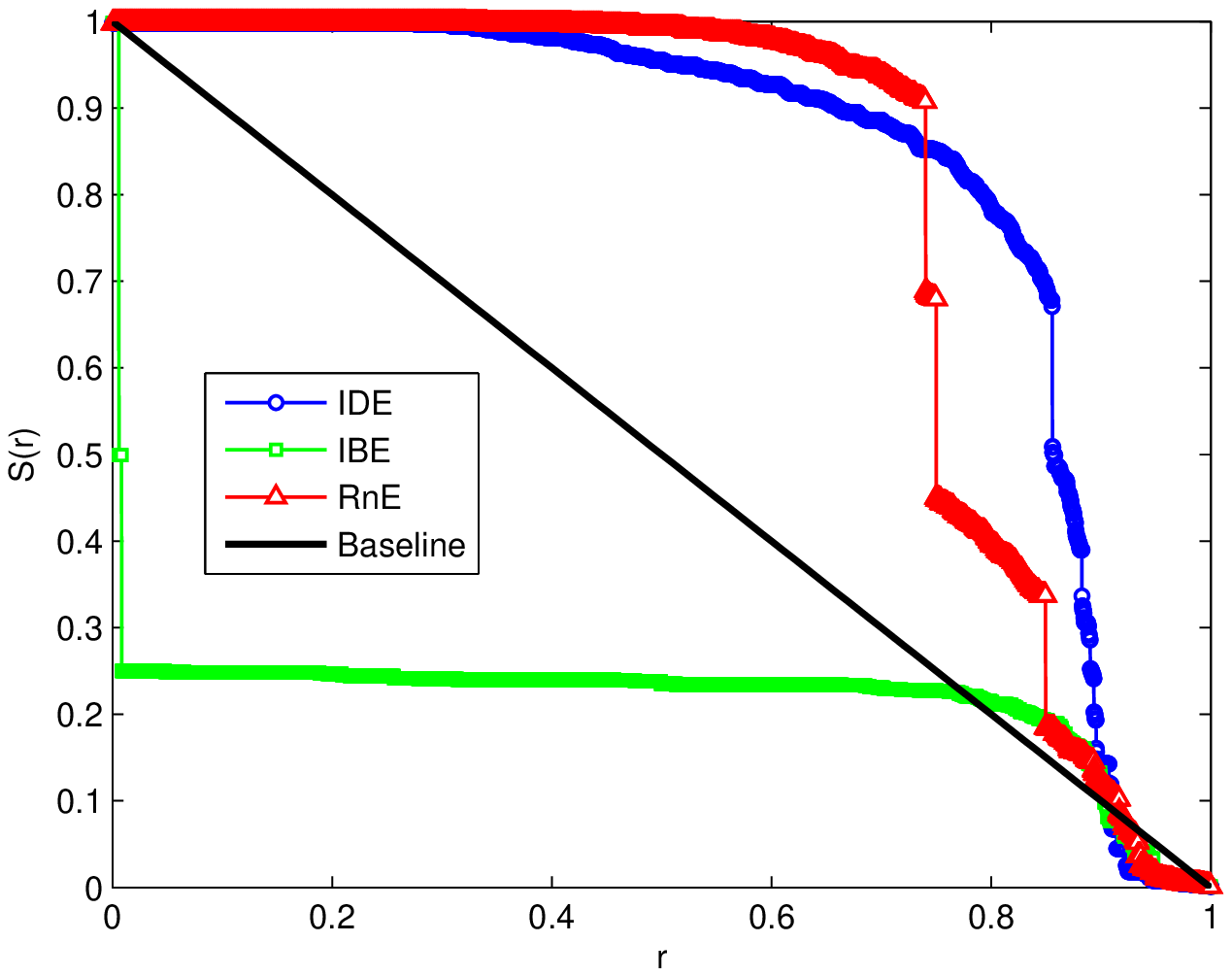}}
    \hspace{-0.7cm}
  \subfigure[Polbook]{
    \label{fig:subfig:c} 
    \includegraphics[width=3.6cm,height=3cm]{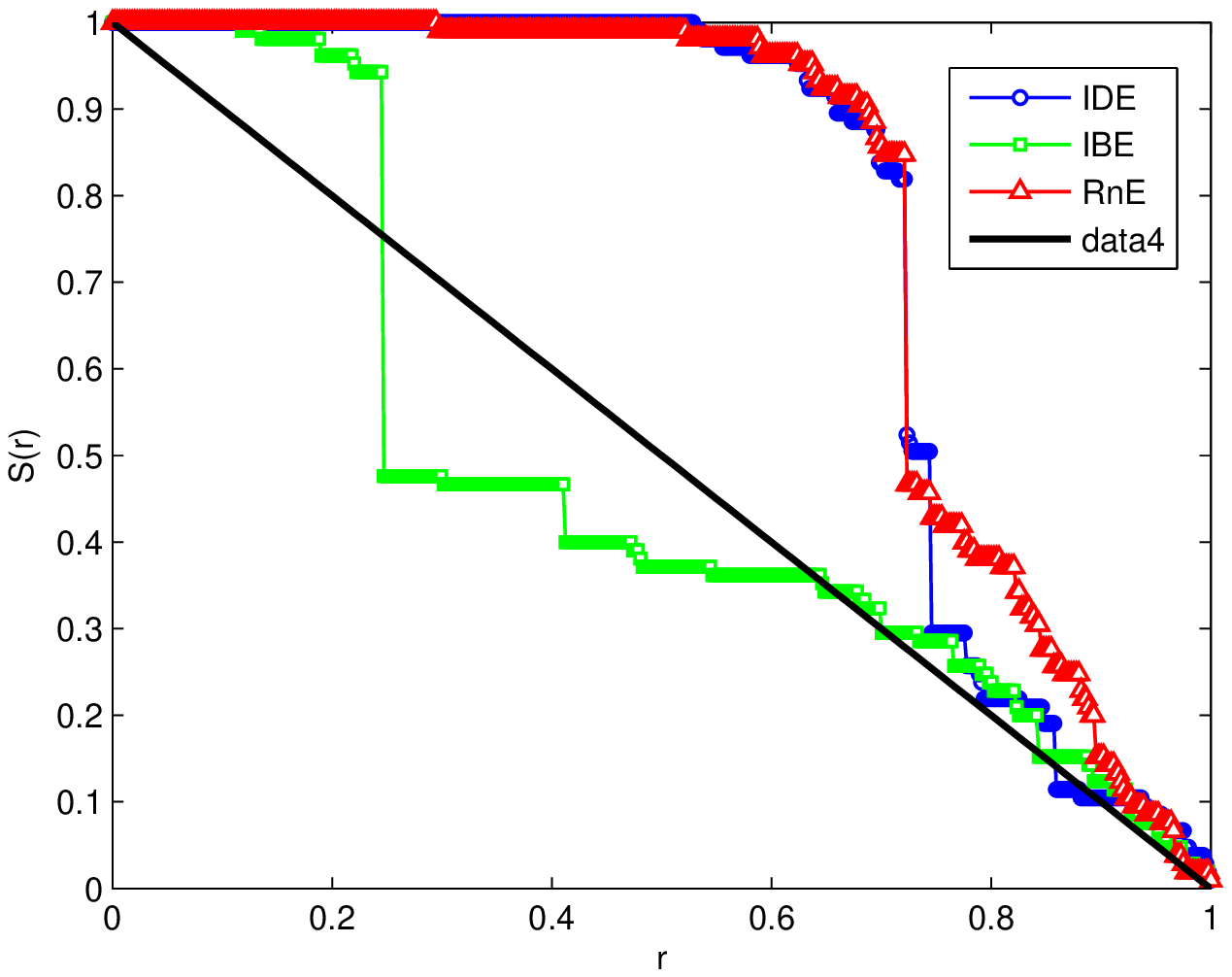}}
    \hspace{-0.7cm}
  \subfigure[Protein]{
    \label{fig:subfig:d} 
    \includegraphics[width=3.6cm,height=3cm]{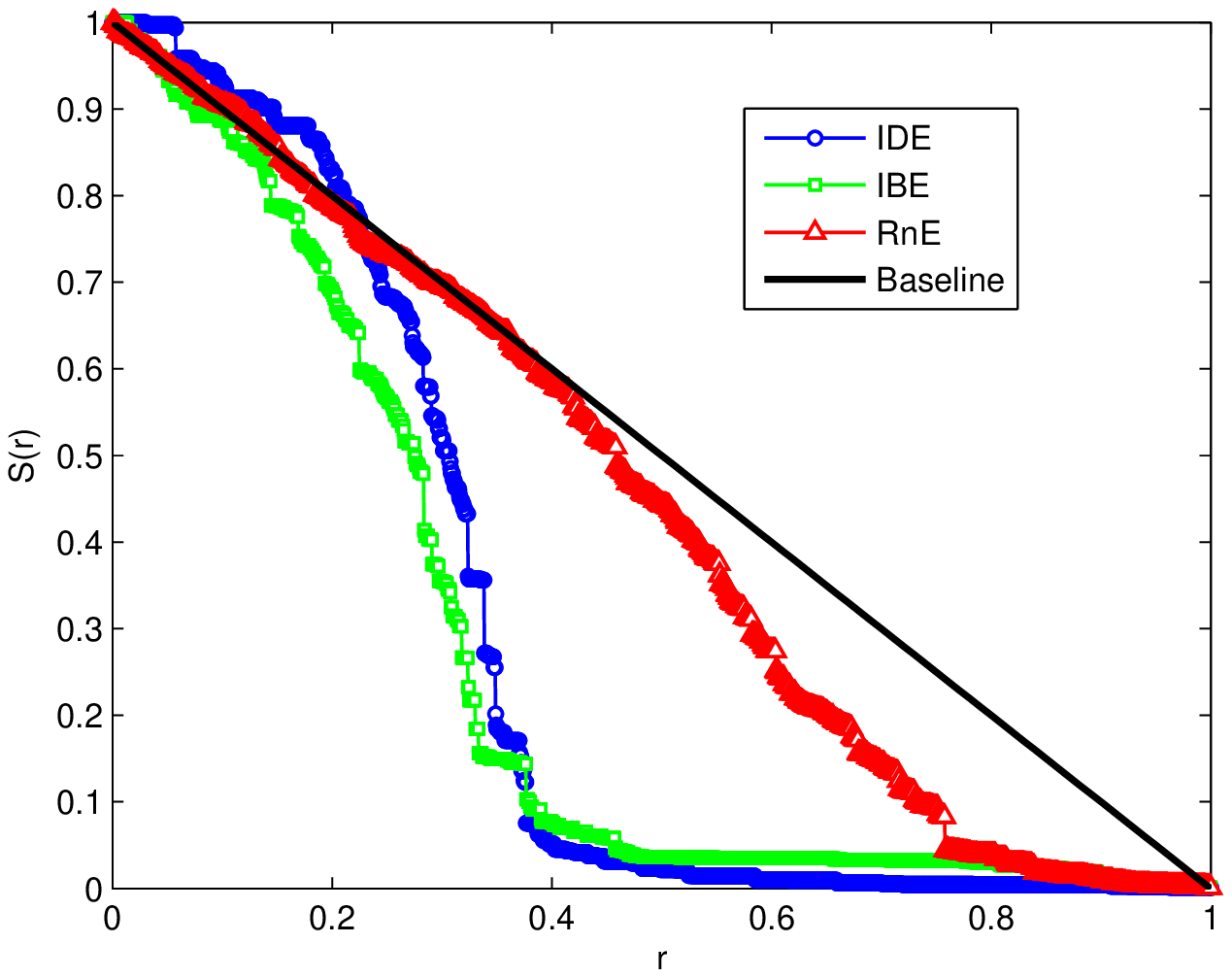}}
   \caption{The results of four selected networks}
  \label{fig:subfig:1} 
\end{figure}

The same as to the definitions in \cite{jun220}, this paper uses the number of the nodes in the giant components as the normalized network performance $s(r)$, which is showed in the vertical axis, and uses the fraction $r$ of removed edges shown in the horizontal axis. Thus, this paper shows the performance curves of the selected networks as Fig. \ref{fig:subfig:1}.

From Fig. \ref{fig:subfig:1}, we can see that the CSF  is the most robust network under three attack strategies; moreover, the majority of the performance curves locate above the baseline, meaning that this network is robust under three attack strategies. Because the CSFN and the Polbook have the community structure, they can not resist the IBE attack strategies, i.e., the corresponding performance curves drops down drastically such that these curves locate under the baseline, however, they are quite robust under the RnE and IDE attack strategies. The Protein is the most fragile network because the majority of the curves locate under the baseline.

\begin{figure}[htbp]
  \subfigure[CSF']{
    \label{fig:subfig:a2} 
    \includegraphics[width=3.6cm,height=3cm]{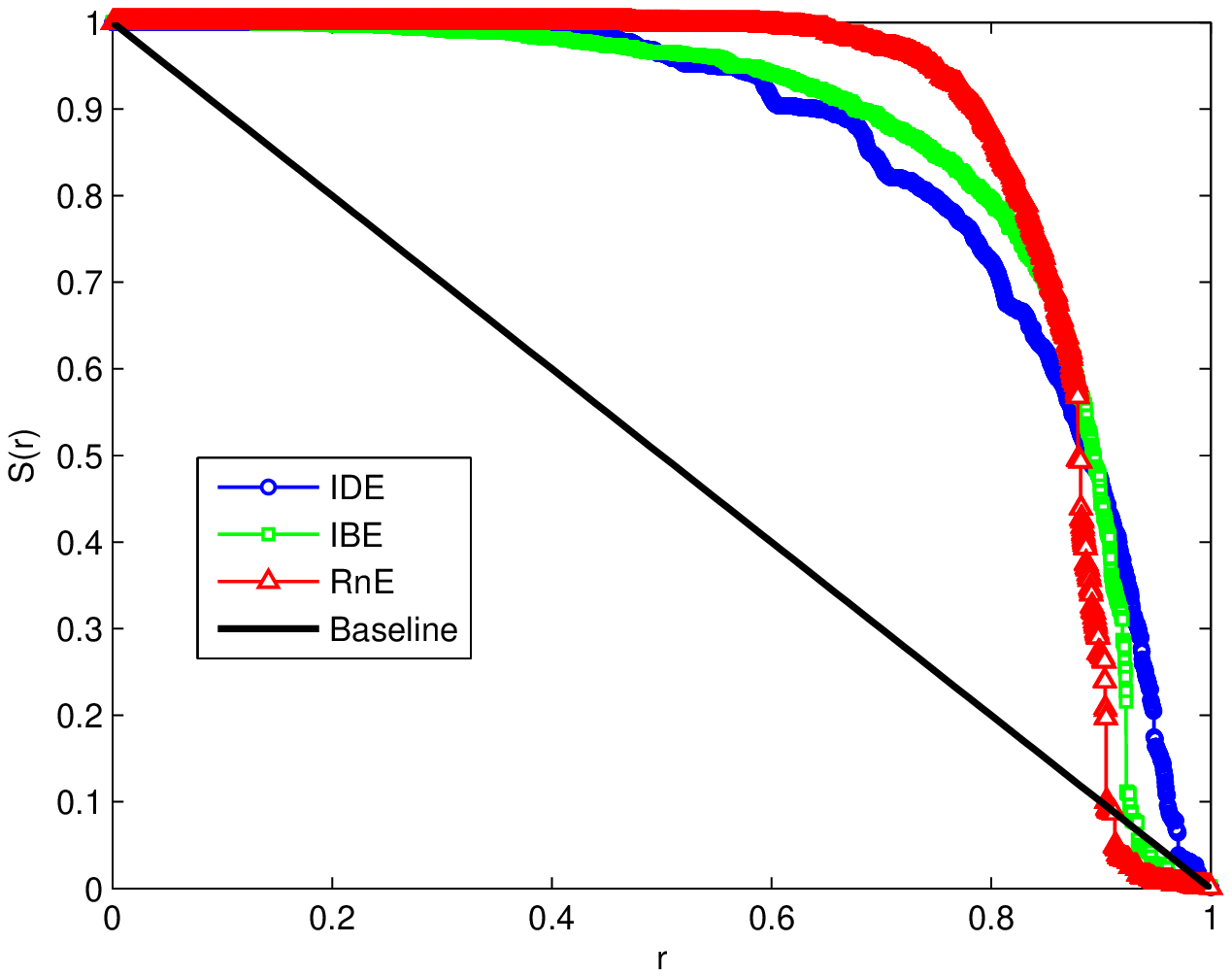}}
  \hspace{-0.7cm}
  \subfigure[CSFN']{
    \label{fig:subfig:b2} 
    \includegraphics[width=3.6cm,height=3cm]{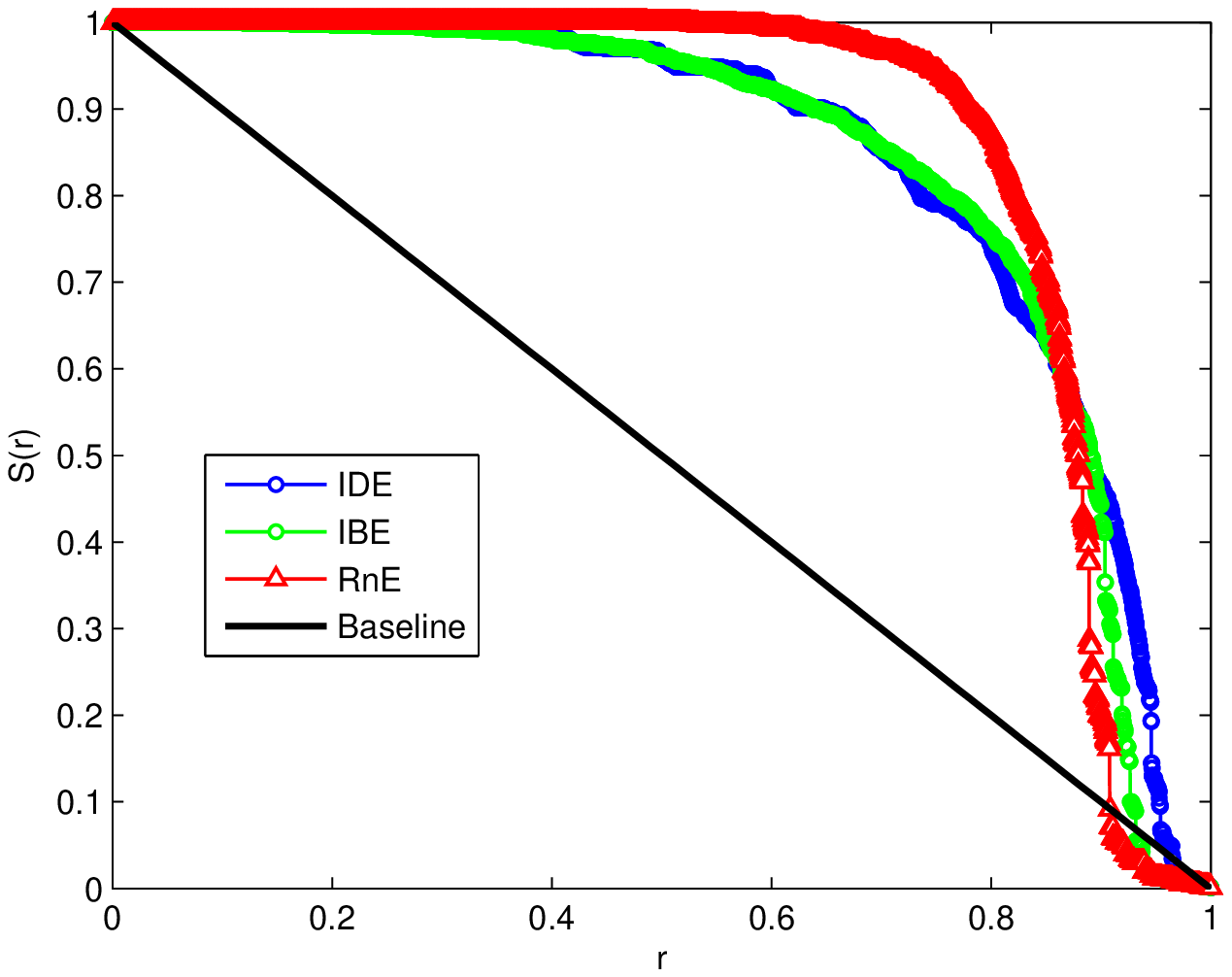}}
    \hspace{-0.7cm}
  \subfigure[Polbook']{
    \label{fig:subfig:c2} 
    \includegraphics[width=3.6cm,height=3cm]{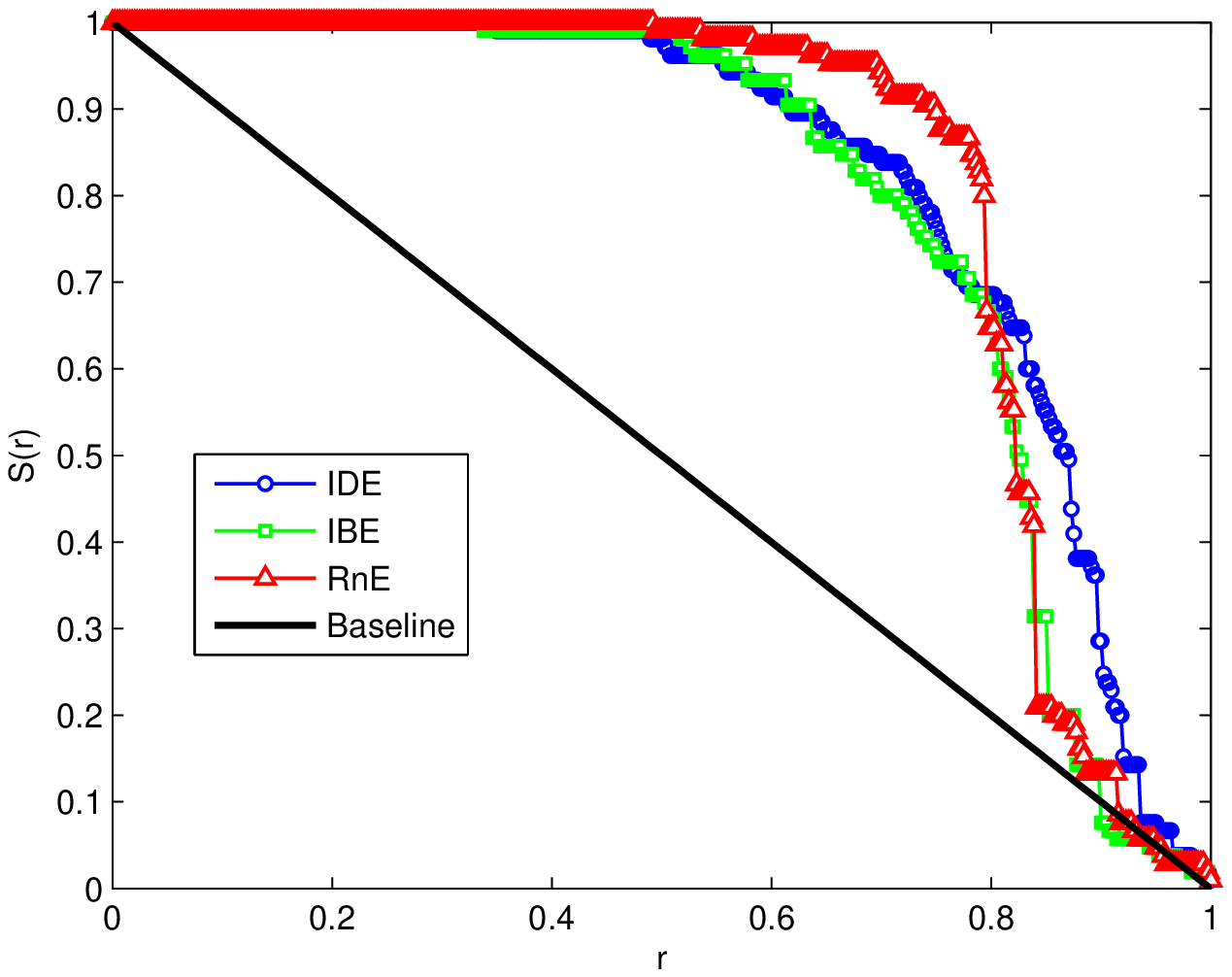}}
    \hspace{-0.7cm}
  \subfigure[Protein']{
    \label{fig:subfig:d2} 
    \includegraphics[width=3.6cm,height=3cm]{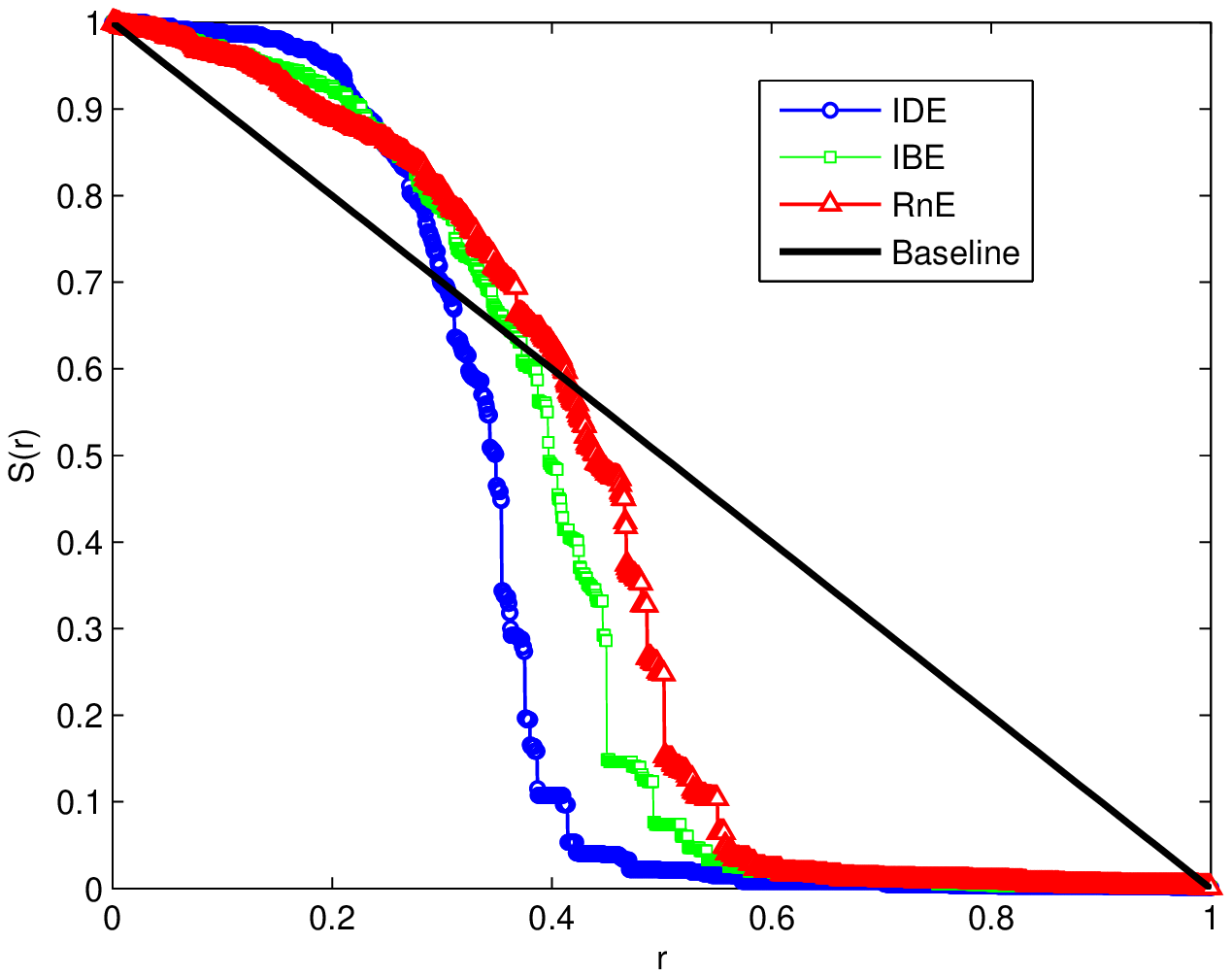}}
   \caption{The results of four random networks responding to the selected networks}
  \label{fig:subfig:2} 
\end{figure}

This paper also uses the same attack strategies to the control group. We plot the performance curves of the corresponding random networks as Fig. \ref{fig:subfig:2}.

From Fig. \ref{fig:subfig:2} we can see that the curves of CSF' is quite similar to that of CSF, meaning that both are quite robust. Moreover, the CSFN' and the Polbook' are similar to the CSFN and the Polbook under the IDE and RnE attack strategies, although the curves under the IBE attack strategies are not. As to the Protein', the majority of the curves locate under the baseline, meaning that this network is still quite fragile under three attack strategies.

Notice that all the networks in the control group are the random networks, the experimental results show that some random networks are robust, and some are fragile. That is to say, random networks can not be regarded as naturally robust. Because the only difference is the average node degree among all these random networks, we attribute the robustness of the random networks to the average node degree. That is, the average node degree determines the robustness or fragility of the random networks.

As to the experimental group, the robustness of the scale-free networks depends on not only the average node degree, but also the compactness. Comparing Fig. \ref{fig:subfig:d} with Fig. \ref{fig:subfig:a}, we can see the impact of the average node degree. Comparing Fig. \ref{fig:subfig:b} with Fig. \ref{fig:subfig:a}, we can see the impact of the compactness. Although the CSF and the CSFN have approximative average node degree, the CSFN can not resist the selective betweenness attacks.

Comparing the experimental group with the control group, we can find that three scale-free networks are much more fragile than the corresponding random networks under certain attack strategies, but the CSF has an approximative robustness to the corresponding random network under all attack strategies.

To quantitatively demonstrate the results, this paper lists the values of the $I$ index as Table \ref{tab2}.

\begin{table}[htbp]\caption{The Invulnerability Index of the networks}
\centering
\begin{tabular}
{|p{54pt}|p{54pt}|p{40pt}|p{40pt}|p{40pt}|p{40pt}|}
\hline
\raisebox{-1.50ex}[0cm][0cm]{Network}&
\raisebox{-1.50ex}[0cm][0cm]{Strategy}&
\multicolumn{4}{|p{160pt}|}{Invulnerability Index}  \\
\cline{3-6}
 &
 &
I$_{0.2}$&
I$_{0.5}$&
I$_{0.7}$&
I$_{1}$ \\
\hline
\raisebox{-3.00ex}[0cm][0cm]{CSF \par }&
IDE& 0.0199 & 0.1203 & 0.2264 & 0.3385  \\
\cline{2-6} & IBE& 0.0200 & 0.1230 & 0.2375 & 0.3592  \\
\cline{2-6} & RnE& 0.0200 & 0.1243 & 0.2390 & 0.3594  \\
\hline
\raisebox{-3.00ex}[0cm][0cm]{CSF' \par }&
IDE&
0.0200 &
0.1225 &
0.2262 &
0.3408  \\
\cline{2-6}
 &
IBE&
0.0198 &
0.1205 &
0.2279 &
0.3452  \\
\cline{2-6}
 &
RnE&
0.0200 &
0.1250 &
0.2435 &
0.3621  \\
\hline
\raisebox{-3.00ex}[0cm][0cm]{CSFN \par }&
IDE&
0.0196 &
0.1202 &
0.2249 &
0.3245  \\
\cline{2-6}
 &
IBE&
0.1252 &
-0.2479 &
-0.2811 &
-0.2817  \\
\cline{2-6}
 &
RnE&
0.0197 &
0.1237 &
0.2431 &
0.3596  \\
\hline
\raisebox{-3.00ex}[0cm][0cm]{CSFN' \par }&
IDE&
0.0200 &
0.1219 &
0.2259 &
0.3405  \\
\cline{2-6}
 &
IBE&
0.0199 &
0.1207 &
0.2243 &
0.3322  \\
\cline{2-6}
 &
RnE&
0.0200 &
0.1250 &
0.2384 &
0.2886  \\
\hline
\raisebox{-3.00ex}[0cm][0cm]{PolBook \par }&
IDE&
0.0184 &
0.1250 &
0.2346 &
0.2595  \\
\cline{2-6}
 &
IBE&
0.0169 &
-0.0503 &
-0.0790 &
-0.1160  \\
\cline{2-6}
 &
RnE&
0.0184 &
0.1231 &
0.2337 &
0.2772  \\
\hline
\raisebox{-3.00ex}[0cm][0cm]{PolBook' \par }&
IDE&
0.0184 &
0.1234 &
0.2261 &
0.3208  \\
\cline{2-6}
 &
IBE&
0.0184 &
0.1234 &
0.2259 &
0.2928  \\
\cline{2-6}
 &
RnE&
0.0184 &
0.1249 &
0.2388 &
0.3198  \\
\hline
\raisebox{-3.00ex}[0cm][0cm]{Protein \par }&
IDE&
0.0066 &
-0.0903 &
-0.1679 &
-0.2118  \\
\cline{2-6}
 &
IBE&
-0.0062 &
-0.1228 &
-0.1959 &
-0.2348  \\
\cline{2-6}
 &
RnE&
0.0002 &
-0.0053 &
-0.0791 &
-0.0634  \\
\hline
\raisebox{-3.00ex}[0cm][0cm]{Protein' \par }&
IDE&
0.0170 &
-0.0463 &
-0.1239 &
-0.1680  \\
\cline{2-6}
 &
IBE&
0.0134 &
-0.0030 &
-0.0775 &
-0.1213  \\
\cline{2-6}
 &
RnE&
0.0114 &
0.0198 &
-0.0504 &
-0.0938  \\
\hline
\end{tabular}
\label{tab2}
\end{table}

From Table \ref{tab2}, we can see that the $I$ index values of the CSF approximates those of the CSF'. The $I_1$ values of the Protein and Protein' are negative, meaning that both are fragile. However, $I_1$ of the Protein' is larger than $I_1$ of the Protein. Moreover, the CSFN is the most fragile network under the selective betweenness attacks.

\section{Discussion}

The experiments have shown some interesting results.

First, previous studies concluded that the random networks are robust, however, this paper discovers that the robustness or fragility of the random networks under the edge attacks are determined by the average node degree. That is, according to different average node degree, some random network are robust under the edge attacks, and some are fragile.

Second, previous studies concluded that scale-free networks are fragile under the selective edge attacks referring to the robustness of the random networks. Because the robustness of the random networks are uncertain when regardless of the average node degree, the conclusions on the scale-free networks should be uncertain. This paper proves that the performance curves of some scale-free networks approximates those of the random networks and the majority of the curves locate above the baseline, therefore, they should be robust. Furthermore, the robustness of scale-free networks under edge attacks depends on the average node degree and the compactness.

Third, the studies on the node attacks relate to the studies on the edge attacks. The node attack strategies can be transferred into the edge attack strategies, meanwhile, some edge attacks strategies can also be transferred into the node attack strategies. Therefore, the conclusions on the edge attacks must be compatible of the conclusions on the node attacks\cite{jun220}. Here we have illustrated that some scale-free networks are robust under selective edge attacks. Noticing that the employed selective edge attacks can be transferred into the node attacks, hence some scale-free networks should be robust under selective node attacks.
Zheng et al. have proved that the scale-free networks would be fragile under the selective node attacks without the consideration of the attack costs, but could be robust with the consideration of the attack costs\cite{152}. That is, the consideration of the attack costs affects the invulnerability of the scale-free networks under the node attacks. Their conclusions are similar to the conclusions in this paper, actually, the attack costs can be regarded as a functions of the edge degree. Therefore, previous studies on the node attacks can be reasonably explained. Above all, the invulnerability of scale-free networks are not determined only by the scale-free property.

\section{Conclusion}

This paper quantitatively analyzed the robustness of the scale-free networks under the edge attacks. The experimental results showed some random networks are fragile under the edge attacks, therefore they can not be taken for granted robust. Because random networks are not certainly robust, the robustness of scale-free networks are uncertain. The experimental results also show that some scale-free networks are fragile under selective edge attacks, similar to previous studies, but some are robust, different to previous studies. By analysis, this paper indicated that the invulnerability of random networks depends on the average node degree; as to scale-free networks, not only the average node degree, but also the compactness are main factors. Finally, this paper reasonably explained previous studies on node and edge attacks.

\section{Acknowledgments}
JQ is grateful for support from the Fundamental Research Funds for the Central Universities (No. CZY12032) and Nature Sience Foundation in Hubei (No.BZY11010). BZ is grateful for support from the State Key Laboratory of Networking and Switching Technology (No. SKLNST-2010-1-04) and the National Natural Science Foundation of China (No.61273213) and (No.60803095).



\end{document}